# Evaluation of Accounting and Market Performance: A Study on Listed Islamic Banks of Bangladesh

- Nusrat Jahan[*]
- M. Ayub Islam[*]

**Abstract**

 This study compared accounting performance of Islamic banks with their market performance and also assessed the effect of firm-specific determinants and cross-sectional effect on accounting and market performance. This study selected all six listed Islamic banks of Chittagong Stock Exchange and the data were collected for the period of 2009 to 2013. This study reported that Social Islamic Bank Limited exhibits superior accounting performance whereas Islami Bank Bangladesh Limited holds better market performance. However, banks exhibiting superior accounting performance reported to have inferior market performance. Further, random-effect model for ROA reports that there exist significant entity or cross-sectional effect on ROA; and operational efficiency and bank size are significantly negatively associated with ROA. However, random-effect model for Tobin's Q failed to ascertain entity or cross-sectional effect on Tobin's Q and also reveals that firm-specific determinants have no significant impact on Tobin's Q.

## 1. Introduction

Banks are financial institution that acts as financial intermediaries to pool financial resources from surplus units and allocate them to deficit units for investment purposes which ultimately results in economic growth of a country. Rapid financial deregulation, consolidation, technological advances and financial innovation are forces that lead to the development of new financial product or instrument or an entirely new financial intermediary system. This development is visible as Islamic banking is fast becoming a widely accepted alternative mode of banking system in the global banking industry. Islamic Banking system has also been playing a crucial role in mobilizing deposits and financing key sectors of the economy in Bangladesh since its inception in 1983.

---

[*]The authors are Assistant Professor, School of Business, Chittagong Independent University, Chittagong and Professor, Department of Accounting and Information Systems, Chittagong University, Chittagong. The views expressed in this article are the authors' own.

At the end of the year 2013, out of 56 commercial banks in Bangladesh, 8 Private Commercial Banks operated as full-fledged Islamic banks, and 16 conventional banks including 3 Foreign Commercial Banks were involved in Islamic banking through Islamic banking branches. The Islamic banking industry continued to show strong growth since its inception as reflected by the increased market share of the Islamic banking industry in terms of assets, financing and deposits of the total banking system. Against this backdrop, application of different measures of evaluation will help examining the profitability and market performance of Islamic banking system in the banking sector of Bangladesh.

## 2. Problem Statement

Performance for a business firm usually refers to the stock price development, profitability and current valuation (Melvin and Hirt, 2005). Thus, performance is a proxy indicator to determine a firm's financial or market related performance, which is mostly measured by non-frontier based financial ratios such as profitability ratio and price to book ratio. Typically, bank performance maybe defined as the reflection of the bank resources used in order to achieve its objectives. The current study evaluates the performance of Islamic banks of Bangladesh by simultaneously applying both accounting-based and market-based measures of performance. The growth of Islamic banking system in the financial sector of Bangladesh motivated the researcher to carry out this research. Further, literature review also reveals that there are no studies till date which evaluated and compared both accounting and market performance of commercial banks in Bangladesh and also examined the equivalency of both measures of performance. Henceforth, this research would be undertaken to fill in these research gaps and thereby contribute to the existing pool of literature on bank performance. It has been conferred in literature review that there are some internal factors that affect the accounting and market performance of commercial banks such as the bank's size, assets management, leverage ratio, operational efficiency ratio, portfolio composition, and credit risk (Almazari, 2011). Most literature on banking has expressed that bank-specific factors originate from banks financial statements and external determinants reflect the economic and legal factors that affect the operation and performance of financial institutions. This study will also assess the impact of selected firm-specific factors such as size, operational efficiency, asset utilization and credit risk on both the accounting and market performance of Islamic banks of Bangladesh. Though the determinants of bank's performance have been well explored in different literatures but there are no studies that focus on comparative accounting and market performance of Islamic banks of Bangladesh. This study would fill a void in the banking performance literature by assessing whether the Islamic banks which are reported to be doing better in terms of financial performance also reported to have comparable market performance.

## 3. Measurement of Firm's Performance

The firm's success is basically explained by its performance over a certain period of time and employing the appropriate criterion of evaluation enables the comparison of firm's performance over different time periods and also within the industry. Researchers have been investigating to determine the measures of performance that can encompass all aspects of performance of a firm. However, no specific criterion with the ability to measure every financial aspect of an organization has been proposed till date. Measurement of performance can offer significant invaluable information to allow management to monitor performance, report progress, improve motivation and communication and pinpoint problems (Waggoner et al. 1999). Further, it is in the firm's best interest to evaluate its performance over time or with others in the industry. Although there are wide varieties of evaluation criterion brought forward by past researches to assess the financial performance of a firm, however, in this study, measurement of performance evaluation are categorized into accounting-based and market-based performance criterion.

### 3.1 Accounting-Based Performance Measurement

Accounting-based measures of performance focus most commonly on company's profitability. The financial ratios including return on assets (ROA), return on equity (ROE) and net interest margin (NIM) are commonly used as accounting-based performance indicators to evaluate profitability condition of commercial banks. According to Hutchinson and Gul (2004) and Mashayekhi and Bazazb (2008), accounting-based performance measures reports the outcome of management actions and hence preferred over market-based measures when the relationship between corporate governance and firm performance is investigated. As a result, when a company is showing a positive performance through ROA, it indicates its achievement of prior planned earnings target (Nuryanah and Islam, 2011). Further, maximization of profit is a short-run goal of a firm and firms are in reality keener to meeting its short-term earnings target than long-run goal of shareholders wealth maximization. The return on asset (ROA) is a substantial performance measure because it is directly related to the profitability of banks (Kosmidou, 2008; Sufian and Habibullah, 2009). ROA measures the profit earned per dollar of assets and reflect how well bank management uses the bank's real investments and resources to generate profits (Ben Naceur, 2003). The higher the value of ROA, the greater is the profitability of banks. Hence, this study employs ROA as an accounting-based performance criterion to evaluate the performance of Islamic banks of Bangladesh.

**3.2 Market-Based Performance Measurement**

A firm's long-term financial goal is creation of wealth for its shareholders through maximizing the market price of its shares; and successfully meeting its short-run earning goals will eventually lead to achieving the long-run financial goal of a firm. Hence, the second type of performance measurement is in focus, in this study, is the market-based indicators which are generally Tobin's Q, market value added, market value to book value and stock return. The market-based measurement is characterized by its forward-looking aspect and its reflection regarding the expectations of the shareholders concerning the firm's future performance, which has its basis on previous or current performance (Wahla et al. 2012; Shan and McIver 2011; Ganguli and Agrawal 2009). Tobin's Q refers to a traditional measure of expected long-run firm performance (Bozec et al. 2010). The employment of market value of equity may reflect the firm's future growth opportunities which could stem from factors exogenous to managerial decisions (Shan and McIver 2011). In addition, a high Q ratio shows success in a way that the firm has leveraged its investment to develop the company, which is valued more, in terms of its market-value compared to its book-value (Kapopoulos and Lazaretou 2007). Chunhachinda and Jumreornvong (1999) used the Tobin's Q to measure the competitiveness of banks and finance companies in Thailand over the period 1990 to 1996. Choi and Hasan (2005) employed the annual stock return and the standard deviation of the daily stock returns to measure the market based performance of Korean commercial bank over the period 1998 to 2002. Jonghe and Vennet (2008) appied the Tobin's Q to measure the European banks' franchise value. Chunhachinda and Li (2011) employed Tobin's Q to measure and compare the competitiveness of Asian banks after recovering from the 1997 financial crisis. Jones et al. (2011) utilize the Tobin's Q to proxy for the charter value in the banking industry. Therefore, this study also employs Tobin's Q as a market-based performance criterion to evaluate the performance of Islamic banks of Bangladesh.

**3.3 Review of Literature on Performance Evaluation of Commercial Banks**

There are a large number of literatures that have evaluated the performance of commercial banks of Bangladesh from different perspectives and several studies also examined the determinants of such performance measures. The following discussion lists few researches that were aimed at evaluating the performance of commercial banks of Bangladesh.

Hassan (1999) examined the performance of Islamic Bank Bangladesh Limited and compared that with other private banks in Bangladesh between 1993 and 1994. While the duration of study was short, the result revealed that in terms of deposits growth and investments growth, performance of Islamic Bank Bangladesh Limited was better than performance of private commercial banks. Apart from that, the researcher found that the key Islamic financial products, mudharabah and musyarakah were not well developed. Siddique and Islam (2001) undertook a study on commercial banks of Bangladesh for the financial year 1980 to1995. The study revealed that in every aspect, Trans National Banks had a commendable performance. But comparing among other groups of banks which are Nationalized Commercial Banks (NCBs), Specialized Banks (SPBs), Private Commercial Banks (PCBs), PCBs had preferred achievement over others aiming profit. On the other hand, Specialized Banks in Bangladesh had a very poor performance. This meager activity affected the overall banking sector's performance. Chowdhury (2002) in his study emphasized that performance of banks requires knowledge about the profitability and the relationships between variables like market size, bank's risk and bank's market size with profitability. The study concluded that the banking industry in Bangladesh is experiencing a major transition for the last two decades. The author recommended that the banks that endure the pressure arising from both internal and external factors prove to be profitable. Hasan and Omar (2006) in their study made a comparative performance analysis between state-owned and privately-owned commercial banks of Bangladesh over the period between 2006 and 2010. ROA and ROE were used to measure profitability and Net profit and net asset efficiencies relative to total employment and total number of branches are used to measure operating efficiency. The results suggest that state-owned banks are as efficient as private banks but private banks have much higher mean values relative to public bank.

Jahangir et al. (2007) stated that the traditional measure of profitability through stockholder's equity is quite different in banking industry from any other sector of business, where loan-to-deposit ratio works as a very good indicator of banks' profitability as it depicts the status of asset-liability management of banks. But banks' market size and market concentration index along with return to equity and loan-to-deposit ratio grab the attention while analyzing the banks' profitability. Chowdhury and Islam (2007) stated that deposit, and loans and advances of nationalized commercial banks (NCBS) are less sensitive to interest changes than those of specialized commercial banks (SCBs). They also suggest that higher return on equity (ROE) is noticeable as it is the primary indicator of bank's profitability and financial efficiency. Nimalathasan (2008) undertook a comparative study of financial performance of banking sector in Bangladesh using CAMELS rating system. The study was done on 6562 Branches of 48 Banks in Bangladesh for the financial year 1999 to 2006. The study revealed that out of 48 banks, 3 banks were rated 01 or Strong, 31 banks were rated 02 or satisfactory, 7 banks were rated 03 or Fair, 5 banks were rated 04 or Marginal and 2 banks obtained 05 or unsatisfactorily rating. 1 Nationalized Commercial Bank (NCB) had unsatisfactorily rating and other 3 NCBs had marginal rating. Chowdhury and Ahmed (2009) in their paper investigated the performance of private commercial banks and revealed that all the commercial banks are able to achieve a stable growth of branches, employees, deposits, loans and advances, net income, earnings per share during the period of 2002 to 2006. Rushdi (2009) in his study compared the performance of Islamic Bank Bangladesh Limited with Janata bank Limited in terms of accounting profitability, partial productivity and total factor productivity over the period from 1983 to 2006. The study confirms that the IBBL performed excellently in terms of labor and capital productivity and TFP over the study period. Sufian and Habibullah (2009) reported in their study that bank specific characteristics, in particular, loan intensity, credit risk and cost have positive and significant impacts on profitability of Bangladeshi banks, while non-interest income exhibits negative relationship with bank profitability. This study found that size has a negative impact on return on average equity (ROAE) while it has positive impact on return on average assets (ROAA) and net interest margin (NIM). Safiullah (2010) in his study emphasized on the financial performance analysis of Conventional and Islamic banks to measure their superiority. The research result based on commitment to economy and community, productivity and efficiency, signifies that interest-based conventional banks are doing better performance than interest-free Islamic banks. But performance of interest-free Islamic banks in business development, profitability, liquidity and solvency is superior to that of interest-based conventional banks. Sarker and Saha (2011) investigated the performance of NCBs, PCBs, FCBs and SCBs through highlighting their profitability, branch productivity, employee productivity and overall productivity and also by using SWOT mix during the period of 2000 to 2009. Sufian and Kamarudin (2012) identified bank specific characteristics and macroeconomic determinants of profitability of 31 commercial banks over the period of 2000 to 2010. The study bring out five bank specific determinants that are important in influencing profitability which are capitalization, non-traditional activities, liquidity, management quality, and size of the bank. Besides, this study found, three macroeconomic determinants significantly influence profitability including growth in GDP, inflation and concentration.

Jahan (2012) evaluated randomly selected six commercial banks of Bangladesh by using widely used indicators of banks' profitability, which are ROA, ROE and ROD. This study investigated the impact of efficiency ratio, asset utilization ratio, asset size and ROD as a determinant of banks' profitability measured by ROA. The results of regression analysis found that operational efficiency, asset size and ROD to be positively related and asset utilization to be negatively related to ROA, but these associations are statistically insignificant. Haque (2013) investigated the financial performance of five private commercial banks in Bangladesh for the period 2006 to 2011 under four dimensions: (1) profitability (2) liquidity (3) credit risk and (4) efficiency. The study concluded that there is no specific relationship between the generation of banks and its performance. The performances of banks are dependent more on the management's ability to formulate strategic plans and the efficient implementation of its strategies.

The above review of past literatures indicates that a comparative study between accounting performance and market performance of commercial banks are yet to be taken in the context of Bangladesh. Hence, the novel feature of current study is expected to broaden the scope of performance evaluation of commercial banks of Bangladesh by shedding light into this less researched area of performance evaluation.

## 4. Research Objectives

The broad objective of this study is to investigate the accounting performance, market performance of Islamic banks of Bangladesh and also to examine the impact of firm-specific factors on performance of Islamic banks.

The specific objectives of this study are as follows:

1) To evaluate and compare the accounting performance of Islamic banks with selected sample banks average.
2) To evaluate and compare the market performance of Islamic banks with selected sample banks average.
3) To examine whether measures of accounting and market performance of selected Islamic banks generate comparable results.
4) To assess the extent to which observed variations in accounting performance of selected Islamic banks are explained by firm-specific factors.
5) To assess the extent to which observed variations in market performance of selected Islamic banks are explained by firm-specific factors.
6) To investigate the effect of cross-sectional differences on ROA and Tobin's Q of selected Islamic Banks.

## 5. Research Methodology
### 5.1 Population, Sample and Sources of Data

This empirical study is based on secondary quantitative data that covers a period of five years from 2009 to 2013. Data required for estimating accounting-based and market-based performance measure and also proxy for selected bank-specific determinants are collected from the annual reports of selected banks. There are 56 commercial banks in Bangladesh, of which 8 are Islamic banks. Among these eight Islamic banks only six are listed in stock exchanges in Bangladesh. Market capitalization data are required for measuring market performance; hence, the sample of this study constitutes all six Islamic banks listed on Chittagong Stock Exchange.

## 5.2 Hypothesis Formulation
The following null hypotheses are developed to fulfill the research objectives:

### 5.2.1 Hypothesis I
The first hypothesis devises the relationship between firm-specific factors which are bank size, operational efficiency, asset utilization and credit risk with accounting performance, measured by ROA. The first null hypothesis is formulated below:

**H1o:** There exists no significant relationship between the firm-specific determinants and ROA of selected Islamic banks.

The first null hypothesis is tested by examining the significance of beta coefficient of random effect model for ROA which is estimated by Generalized Least Squares (GLS) regression estimator. If the calculated probabilities of all beta coefficients of selected determinants are less than 0.05 level of significance, then the first null hypothesis (H1o) will be rejected.

### 5.2.2 Hypothesis II
The second hypothesis formulates the relationship between firm-specific factors which are bank size, operational efficiency, asset utilization and credit risk with market performance, measured by Tobin's Q. The second null hypothesis is formulated below:

**H2o:** There exists no significant relationship between the firm-specific determinants and Tobin's Q of selected Islamic banks.

The second null hypothesis is also tested by examining the significance of beta coefficient of random effect model for Tobin's Q which is estimated by Generalized Least Squares (GLS) regression estimator. If the calculated probabilities of all beta coefficients of selected determinants are less than 0.05 level of significance, then the second null hypothesis (H2o) will be rejected.

### 5.2.3 Hypothesis III and IV

Hypothesis III and IV are extensions of hypothesis I and II respectively, which are formulated to examine whether cross-sectional or entity differences have any influence on dependent variable measured by ROA and Tobin's Q. The third and fourth null hypotheses are formulated below:

**H3o:** There exists no cross-sectional or entity effect on ROA of selected Islamic banks.

**H4o:** There exists no cross-sectional or entity effect on Tobin's Q of selected Islamic banks.

The third and fourth null hypotheses are tested by examining the significance of estimated random effect model for ROA and Tobin's Q as a whole. If the calculated probability of Wald chi-square test is less than 0.05 level of significance, signifying that the estimated model for ROA is statistically significant as a whole, then the third null hypothesis (H3o) will be rejected. If the calculated probability of Wald chi-square test is less than 0.05 level of significance, signifying that the estimated model for Tobin's Q is statistically significant as a whole, then the fourth null hypothesis (H4o) will be rejected. A statistically significant random-effect model for ROA and Tobin's Q would suggest that there exists significant entity or cross-sectional effects on accounting performance and market performance of selected Islamic banks.

### 5.3 Measures of Performance Evaluation and Bank-specific Determinants

In line with existing literature, traditional non-frontier based financial ratio, ROA, would be used to measure accounting performance (Ali et al. 2011) and Tobin's Q would be used to evaluate market performance (Siddique and Shoaib 2011). The proxies or ratios used for measuring dependent and explanatory variables are listed in the following table:

| Dependent Variables | Description | Independent or Explanatory Variables | Description |
|---|---|---|---|
| ROA | Net Income /Total Asset | Bank Size (size) | ln (Total Assets) |
| Tobin's Q | Market Value of Bank /Total Asset | Credit Risk (CR) | Classified Investment /Total Investment |
| | | Operational Efficiency (OE) | Total Operating Expense /Operating Income |
| | | Asset Utilization (AU) | Operating Income / Total Assets |

**Table 1: Summary of Dependent and Independent Variables**

## 5.4 Model Specification

The data collected from sample constitutes a panel database for this study since it includes both cross-sectional and time-series data for six Islamic banks over the period of 2009 to 2013. Hence, panel data model is considered appropriate for this study for investigating the impact of explanatory variables on accounting performance and market performance of Islamic banks. Random-effects regression model is preferred for this study with the assumption that cross-sectional or entity differences may have some influence on dependent variable. Besides, random–effect model allows generalizing the inferences beyond the sample used in the model. A random effect model estimated by Generalized Least Squares (GLS) regression would be used to determine the association of explanatory variables with performance. The panel data model for random-effect estimation is expressed as below:

$$DR_{i,t} = \acute{\alpha} + \beta_1 (Size)_{i,t} + \beta_2 (CR)_{i,t} + \beta_3 (OE)_{i,t} + \beta_4 (AU)_{i,t} + \mathcal{E}_{i,t}$$

Where, Dependent variable: $DR_{i,t}$ = ROA or Tobin's Q of bank i at time t
Independent Variables are as follows:
$(Size)_{i,t}$ = Size of bank i at time t
$(CR)_{i,t}$ = Credit Risk of bank i at time t
$(OE)_{i,t}$ = Operational Efficiency of bank i at time t
$(AU)_{i,t}$ = Asset Utilization of bank i at time t
$\acute{\alpha}$ = is the intercept
$\mathcal{E}_{i,t}$ = is the random error term for firm i at time t

## 5.5 Statistical Tools Used for Analysis

In this study both descriptive and inferential statistics are used and parametric tests are applied for hypothesis testing. Descriptive statistics measures used includes arithmetic mean, minimum, maximum, standard deviation and trend analysis. Evaluation and comparison of accounting and market performance are presented through Histogram, where each column represents a bank, defined by a quantitative variable which are ROA and 'Tobin's Q'. To evaluate accounting performance of each bank, five-year average ROA of each Islamic bank will be compared with the estimated average ROA of all six listed Islamic banks and this comparison is presented through a Histogram. Further, to evaluate market performance of each bank, five-year average 'Tobin's Q' of each bank will be compared with the estimated average 'Tobin's Q' of all six listed Islamic banks and this comparison is also presented through a Histogram. A Line Chart is used to plot five-year average ROA and 'Tobin's Q' of each bank to examine whether accounting and market performance of Islamic banks generate comparable performance result. Inferential statistics are applied with the purpose of generalizing the findings of the sample to the population it represents, and they can be classified as either parametric or non-parametric. Parametric tests make assumptions about the parameters or properties of a population, whereas nonparametric tests do not include such assumptions or include fewer. Parametric inferential statistics used for testing of hypotheses is a panel data model known as Generalized Least Squares (GLS) random effect model. For generating descriptive statistics and conducting panel data analysis, the study uses the statistical software 'STATA'. The charts are created using software MS-Excel.

## 6. Findings and Analysis
### 6.1 Descriptive Statistics of Panel Data

Table 2 reports descriptive statistics of panel data, where total number of banks are n= 6, time T= 5 years and total number of observations are N=30. The overall, between and within value of ROA, Tobin's Q, asset utilization, operational efficiency, credit risk and bank size are calculated over 30 bank-years data.

| Variable | Mean | Std. Dev | Min. | Max. | Obs. |
|---|---|---|---|---|---|
| ROA: Overall | 0.017459 | 0.006895 | 0.0053 | 0.0354 | N = 30 |
| Between | | 0.004204 | 0.01278 | 0.02322 | n = 6 |
| Within | | 0.005684 | 0.00992 | 0.03199 | T = 5 |
| Tobin's Q: Overall | 0.198311 | 0.450993 | 0.0000085 | 2.4 | N = 30 |
| Between | | 0.276028 | 0.000025 | 0.732 | n = 6 |
| Within | | 0.371096 | -0.39369 | 1.86631 | T = 5 |
| Asset Utilization: Overall | 0.046468 | 0.011015 | 0.0267 | 0.07846 | N = 30 |
| Between | | 0.008949 | 0.0291 | 0.05433 | n = 6 |
| Within | | 0.007230 | 0.033239 | 0.07059 | T = 5 |
| Opr. Efficiency: Overall | 0.364743 | 0.072488 | 0.2209 | 0.54 | N = 30 |
| Between | | 0.057469 | 0.30172 | 0.4593 | n = 6 |
| Within | | 0.049064 | 0.2544833 | 0.44838 | T = 5 |
| Credit Risk: Overall | 0.02781 | 0.013068 | 0.0094 | 0.0647 | N = 30 |
| Between | | 0.008371 | 0.0167 | 0.04112 | n = 6 |
| Within | | 0.010505 | 0.00887 | 0.06417 | T = 5 |
| Bank Size: Overall | 6.154747 | 2.214164 | 4.6019 | 11.12 | N = 30 |
| Between | | 2.369443 | 4.9532 | 10.9678 | n = 6 |
| Within | | 0.250284 | 5.469367 | 6.79406 | T = 5 |

**Table 2: Descriptive Statistics of Panel Data**

The overall average ROA is 0.017459 and overall standard deviation of 30 observations is 0.006895. Minimum and maximum statistics reports, overall ROA calculated for 30 bank-years data varied between 0.0053 to 0.0354. The between standard deviation of ROA calculated across six banks is 0.004204. Between standard deviation of ROA calculated for each bank, on an average, varies across bank by 0.01278 to 0.02322. Within standard deviation of ROA is 0.005684, indicating deviation from each bank's five years average and it varies between 0.00992 to 0.03199. The overall standard deviation of ROA calculated for 30 bank-years data is 0.006895. Between standard deviation of ROA indicates that the variation that exists in ROA across banks is 0.004204 and it is almost close to that of observed within a bank over time which is 0.005684. Descriptive statistics also reports overall, within and between mean, maximum, minimum and standard deviation for Tobin's Q, asset utilization, operational efficiency, credit risk and bank size. From panel data set, the random-effect model is generated.

## 6.2 Evaluating and Comparing the Accounting Performance of Selected Banks

Banks' profitability is a vital issue of contemporary banking field that grace its role by emphasizing on the financial soundness of banks. This study assumes that the banks' performance is represented by their ability to generate sustainable profitability as measured in this study by ROA.

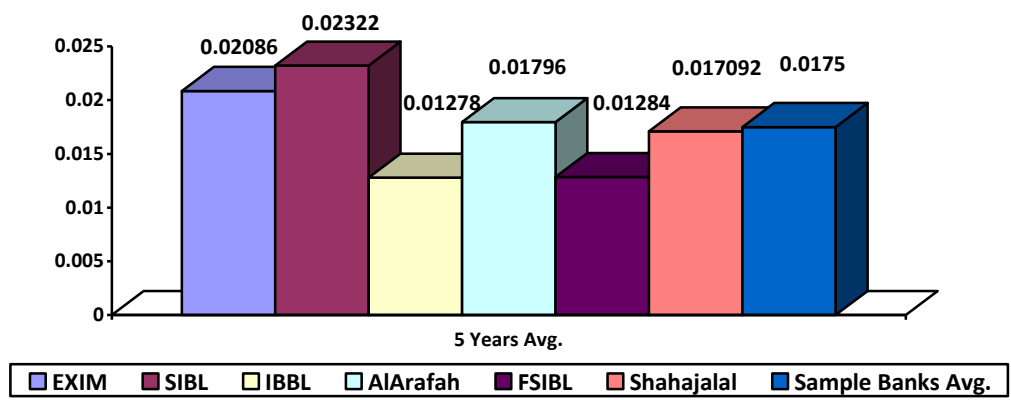

**Chart 1: Comparison of Individual Bank's Average ROA with Six Banks' Average**

Chart 1 reports Return on Asset (ROA), which is an indicator of how profitable a company is, relative to its total assets. ROA is calculated by dividing a company's annual earnings by its total assets and displayed as a percentage. Sometimes, ROA is also referred as "return on investment". ROA gives an idea as to how efficient management is at using its assets to generate earnings. ROA for public limited companies can vary substantially and will be highly dependent on the nature of industry. This is why, when using ROA as a comparative profitability measure, it is best to compare it against the ROA of a similar company. The ROA figure gives investors an idea of how effectively the company is converting the money it has to invest, into net income. The higher the ROA ratio, the better will be the profitability, because the company is earning more money on less investment. The average ROA of all listed Islamic banks set as benchmark in this study is 1.75%. Compared to this benchmark ROA, Social Islami Bank (SIBL) reports average ROA of 2.322%, indicating that this bank is relatively better compared to other five Islamic banks at converting its investment into profit. However, compared to benchmark ROA, the minimum ROA is reported by Islami Bank Bangladesh Limited (IBBL) and First Security Islami Bank Limited (FSIBL), which are 1.278% and 1.284% respectively.

**6.3 Evaluating and Comparing the Market Performance of Selected Banks**

Using Tobin's Q as a measure of market performance, the study seeks to examine the relative performance of selected sample banks. Tobin's Q is the ratio of the market value of a firm to the replacement cost of the firm's assets (Tobin, 1969).

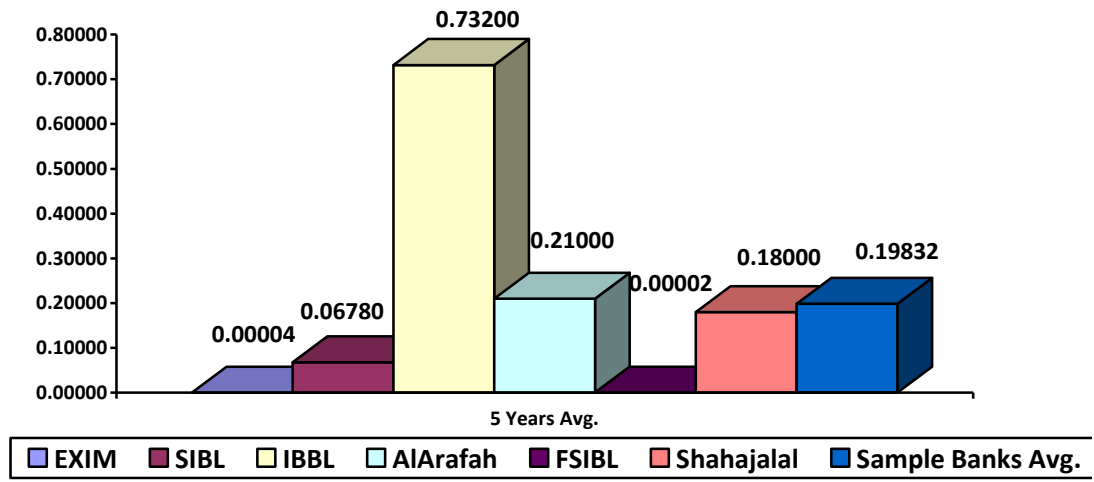

**Chart 2: Comparison of Individual Bank's Average Tobin's Q with Six Banks' Average**

Tobin's Q as a measure of firm's performance is not used as often as either accounting rate of return or price to cost margins (Carlton and Perlof 2005). According to Tobin's Q, if a firm is worth more than its value based on what it would cost to rebuild it, then excess profits are being earned. These profits are above and beyond the level that is necessary to keep the firm in the industry. Tobin's Q ratio is based on the work of James Tobin, who suggested that a fairly priced company ought to have a price equal to its total asset value (Tobin, 1969). Thus, when Tobin's Q ratio is less than one, it means that the market value of the company is less than the total asset value, indicating that it is undervalued. Likewise, when it is more than one, it indicates that the market value is higher than the total asset value and that the company might be overvalued. Tobin's Q ratio is also termed simply a 'Q' ratio. Firms with high Tobin's Q ratio, i.e. greater than one, have been assumed to offer attractive investment opportunities for investors (Lang et al. 1989) and also expected to have higher growth potential (Tobin and Brainard 1963; Tobin 1969) and it indicates that management has been better utilizing the assets of the firm (Lang et al. 1989). In Chart 2, all the banks have Tobin's Q less than one, which indicates that the actual or intrinsic value of the assets of all selected Islamic banks are not properly assumed by the investors and hence all the banks remained undervalued in the market. Further, the estimated benchmark, which is the average Q of six listed Islamic banks calculated as 0.1983, also indicates that the Islamic banks remained undervalued in the stock market during the period of study. However, compared to this estimated yardstick, the relative market performance of Islami Bank Bangladesh Limited (IBBL) and Al-Arafah Islamic Bank are reportedly better.

**6.4 Comparing Accounting Performance with Market Performance of Selected Banks**

Accounting performance and market performance both act as an indicator of a firm's success or failure in a business environment. Therefore, the following Line Chart 3 is used to examine whether the banks, exhibiting superior accounting performance, are also reported to be doing better in the stock market. Therefore, this study reveals whether the results of accounting performance measured by ROA and market performance measured by Tobin's Q are comparable or not.

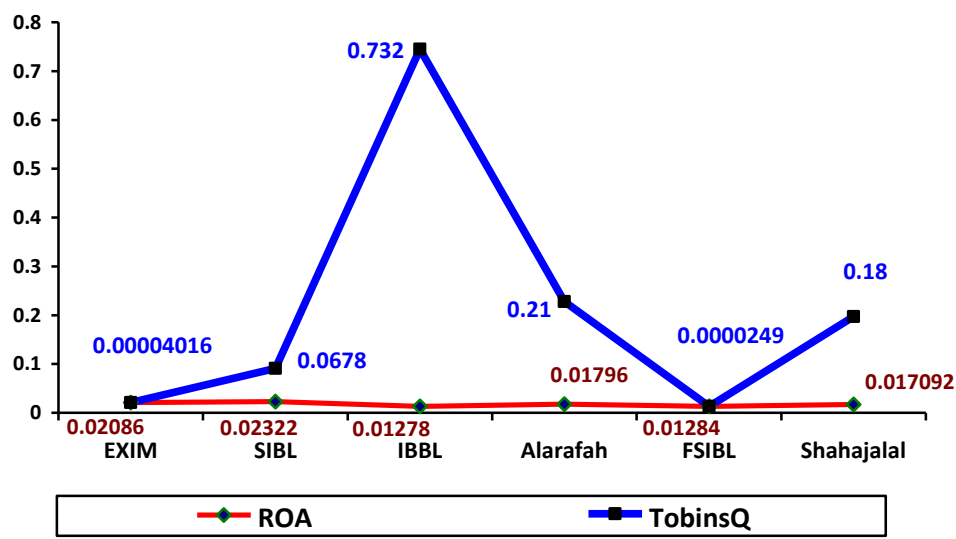

**Chart 3: Comparison of Accounting (ROA) and Market Performance (Tobin's Q)**

The result reported in Chart 3 indicates that though accounting performance of Islami Bank Bangladesh Limited (IBBL) is reported to be inferior having least ROA of 1.278% but its market performance is better compared to six listed banks as reported by highest Tobin's Q score of 0.732. However, superior accounting performance is reported for Social Islami Bank Limited (SIBL) with average ROA of 2.322% but market performance of the same bank reported by Q score of 0.0678 is quite inferior compared to Islami Bank Bangladesh

Limited (IBBL), Shahajalal Islami Bank Limited and Al-Arafah Islami Bank Limited. Comparison depicted in Chart 3 suggest that that since the yardstick of accounting and market performance are not comparable, hence, the superior accounting performance of banks may not necessarily leads to improved market performance.

**6.5 Analysis of Random-Effect Model for ROA and Tobin's Q**

Random-effect model is applied to observe how the variations in accounting performance and market performance of Islamic banks are explained by different firm-specific factors and also to examine whether cross-sectional or entity differences have any influence on dependent variables. The result of random-effects model is presented in Table 3 and Table 4.

| Random-effect GLS regression | Number of observations =30 Number of groups =6 Average observation per group = 5 | | |
|---|---|---|---|
| ROA (dependent variable) | Coefficient | Z-statistics | Probability |
| Asset Utilization | 0.0045563 | 0.04 | 0.969 |
| Operational Efficiency | -0.09221149 | -4.83 | 0.000** |
| Credit Risk | 0.071243 | 1.03 | 0.302 |
| Bank Size | -0.0013725 | -2.25 | 0.024** |
| Constant | 0.057348 | 4.60 | 0.000 |
| R-square: Within=0.7374  Between =0.3636  Overall= 0.5580 | | | |
| Wald Chi-square F(4) = 47.41    Probability =0.000** | | | |
| Notes: ** means statistically significant at 5% level of significance | | | |

**Table 3: Results of Random-Effects Model for ROA**

The result of random-effect regression model for ROA is reported in Table 3. In this table, operational efficiency and bank size are found to be significant explanatory variables of ROA. Table 3 reports that the beta coefficient of bank size is negative and its association with ROA is statistically significant at 5% level of significance. Therefore, 1 unit increase in bank size reduces the ROA by the amount of beta coefficient which is 0.0013725 units. This finding implies that Islamic banks are failing to take advantage of cost reduction that comes along with economies of scale. However, the result of this study shows conformity to prior study by Athanasoglou et al. (2005), which suggest that if the size of bank becomes larger, phenomenon of the diseconomies of scale may appear, as it becomes more difficult for management to conduct surveillance and the higher the level of bureaucracy creates a negative impact on banks profitability. Further, table 3 reports that the beta coefficient of operational efficiency is also negative and its association with ROA is statistically significant at 5% level of significance. Therefore, 1 unit increase in operating efficiency ratio reduces the ROA by the amount of beta coefficient which is 0.09221149 units. High operating efficiency ratio indicates having higher percentage of cost compared to income hence signifying poor expenses management by the bank. This finding is at par with prior studies by Curak et al. (2012), Alper and Anbar (2011), and Almazari (2014) that have reported that there is a negative relation between operating inefficiency and profitability. Hence, this study rejects the first null hypothesis (H1o) and concludes that there exists significant association between the firm-specific determinants and ROA of selected Islamic banks. Wald chi-square test is used to show whether all the coefficients in the random-effect model are different from zero. The estimated probability of Wald chi-square test is less than 0.05, hence the random-effect model for ROA as a whole is found to be statistically significant at 5% level of significance. Hence this study also rejects the third null hypothesis (H3o) and reports that there exists significant entity or cross-sectional effects on ROA by selected Islamic banks.

| Random--effect GLS regression | Number of observations =30 Number of groups =6 Average observation per group = 5 | | |
|---|---|---|---|
| Tobin's Q (dependent variable) | Coefficient | Z-statistics | Probability |
| Asset Utilization | 4.944365 | 0.39 | 0.694 |
| Operational Efficiency | 1.400537 | 0.69 | 0.489 |
| Credit Risk | -9.287875 | -1.24 | 0.216 |
| Bank Size | 0.0253017 | 0.31 | 0.759 |
| Constant | -0.4397097 | -0.32 | 0.746 |
| R-square: Within=0.0648 Between =0.0240 Overall= 0.0491 | | | |
| Wald Chi-square F(4) = 1.62      Probability =0.8049 | | | |
| Notes: ** means statistically significant at 5% level of significance | | | |

**Table 4: Results of Random-Effects Model for Tobin's Q**

The result of random-effect regression model for Tobin's Q is reported in Table 4. Table 4 reports that the beta coefficients of asset utilization, operational efficiency and bank size are positive but beta coefficient of credit risk is negative. However, their association with Tobin's Q is statistically insignificant at 5% level of significance since their p-values are more than 0.05. Hence, this study fails to reject the second null hypothesis (H2o) and concludes that there exist no significant association between the firm-specific determinants and Tobin's Q of selected Islamic banks. Wald chi-square test is used to show whether all the coefficients in the random-effect model are different than zero. The estimated probability of Wald chi-square test is more than 0.05, hence the random-effect model for Tobin's Q as a whole is found to be statistically insignificant at 5% level of significance. Therefore, this study also fails to reject the fourth null hypothesis (H4o) and concludes that there is no significant entity or cross-sectional effects on Tobin's Q by selected Islamic banks.

**7. Conclusion**

The principal aim of this study is to compare accounting performance of Islamic banks with their market performance and also to assess the effect of firm-specific determinants and entity or cross-sectional effect on accounting and market performance. This study selects all six listed Islamic banks of Chittagong Stock Exchange and the data are collected for the period of 2009 to 2013. Current study reveals that, relative to all selected banks, Social Islamic Bank Limited has superior accounting performance in terms of ROA, whereas Islami Bank Bangladesh Limited reports better market performance with Tobin's Q. However, this research also reveals that banks exhibiting superior accounting performance reportedly have inferior market performance. It is reasonable to assume that banks that are able to meet their short-term goals of meeting targeted profit, will eventually be creating wealth for its shareholders by maximizing the market value equity. Despite being profitable and with other fundamentals in place, if a bank's intrinsic value of assets is not reflected on its market value or stock price, this may be due to incorrect valuation of that bank in the stock market. However, the reason behind all selected Islamic banks to be undervalued in the stock market during the period of this study could also be the consequence of investors' lack of confidence on the stability of stock market. Furthermore, this study reports that there exists significant entity or cross-sectional effect on ROA; and operational efficiency and bank size are significant explanatory variables of ROA of selected Islamic banks. This implies size of banks assets and cost control is influential factors in shaping profitability of Islamic banks. The negative association of ROA with operational efficiency is justifiable because it indicates that banks are inefficiently managing expenditures, thus leading to reduction in profitability. The inverse relationship of bank size and ROA may imply that small banks are failing to take benefits arising from economies of scale while growing their business. However, this study fails to ascertain entity or cross-sectional effect on Tobin's Q and also reveals that firm-specific determinants have no significant impact on Tobin's Q of selected Islamic banks. Finally, on the basis of selected banks, this study concludes that the accounting performance and market performance may not necessarily generate comparable results. However, there remains scope for future researches by including all listed commercial banks of Bangladesh to substantiate the outcome of this current study.

The expected contribution of this study to the field of bank management is to assist decision makers in efficient financial resource allocation for Islamic banks and also to pay more attention to the relevant activities that exert potential and strong impact on the both accounting and market performance. This study would also be contributing to the academic field by providing a comprehensive analysis of two methods for evaluating and comparing banking performance and also to fill important gaps in literature mentioned earlier.

Ben Naceur, S. (2003), "The determinants of the Tunisian banking industry profitability: Panel evidence", Paper presented at the Economic Research Forum (ERF) 10th Annual Conference, Marrakesh-Morocco.

Bozec, et al. (2010), "Governance – performance relationship: A Re-examination Using Technical Efficiency Measures", *British Journal of Management*, No. *21*, pp. 684–700.

Chowdhury, A. (2002), "Politics, Society and Financial Sector Reform in Bangladesh", *International Journal of Social Economies*, Vol. 29, No. 12, pp. 963 – 988.

Chowdhury, H. A and Islam M.S. (2007), "Interest sensitivity of loan and advances: A Competitive study between nationalized commercial banks and specialized commercial banks", *ASA University review*, Vol. 1, No. 1, pp. 124-141.

Chunhachinda, P. and Jumreornvong, S. (1999), "Competitiveness of banks and finance companies in Thailand", *Thammasat Review*, Vol. 4, No. 1, pp. 59-88.

Choi, S., and Hasan, I. (2005), "Ownership, governance, and bank performance: Korean experience", *Financial Markets, Institutions and Instruments,* Vol. 14, pp. 215-241.

Chunhachinda, P. and Li, L. (2011), "Competitiveness of Asian banks after recovering from the 1997 financial crisis", Working Paper.

Chowdhury, T.A., and Ahmed, K. (2009), "Performance Evaluation of Selected Private Commercial Banks in Bangladesh", *International Journal of Business Management*, Vol. 4, No. 4.

Curak, et al. (2012), "Profitability determinants of the Macedonian banking sector in changing environment", *Procedia- Social and Behavioral Sciences*, Vol. 44, pp. 406-416.

Carlton, W.D. and Perlof, M. J. (2005), "Modern Industrial Organization. *Industry Structure and Performance"*, Pearson, Cambridge, UK.

Ganguli, S. K., and Agrawal, S. (2009), "Ownership structure and firm performance: An empirical study on listed Mid-Cap Indian Companies".

Haque, S. (2013), "The Performance Analysis of Private Conventional Banks: A Case Study of Bangladesh", *IOSR Journal of Business and Management,* Vol. 12, No. 1, pp. 19-25.